\documentclass[twocolumn,prd,superscriptaddress,amssymb,nofootinbib,amsmath,nobibnotes,aps,prlgroupedaddress]{revtex4}  
\usepackage{graphicx}  
\usepackage{dcolumn}   
\usepackage{bm}        
\usepackage{amssymb}   
\usepackage[]{amsmath,amssymb}
\usepackage{graphics,epsfig}
\usepackage{float}
\usepackage{mathtools}
\usepackage{esint}
\usepackage[utf8]{inputenc}
\usepackage[english]{babel}
\usepackage{caption}
\usepackage{subcaption}
\usepackage{mathtools}
\usepackage{xcolor}
\usepackage{float}  
\usepackage{setspace} 
\usepackage{indentfirst}  
\usepackage{cancel}  
\usepackage{hyperref}  
\hypersetup{
            colorlinks=true,
            linkcolor=black,
            urlcolor=blue,
            } 
\usepackage{tcolorbox}

\begin{document}
  
\title{Ultralight vector dark matter, anisotropies and cosmological adiabatic modes
}
\author{Tomas Ferreira Chase}
\email{tferreirachase@df.uba.ar}
\author{Diana Lopez Nacir}
\email{dnacir@df.uba.ar}
\affiliation{Universidad de Buenos Aires, Facultad de Ciencias Exactas y Naturales, Departamento de Física. Buenos Aires, Argentina.}
\affiliation{CONICET - Universidad de Buenos Aires, Instituto de Física de Buenos Aires (IFIBA). Buenos Aires, Argentina}

\date{\today}

\begin{abstract}
    In this work we study the background evolution and the early time behavior of scalar cosmological perturbations  with an  ultralight vector dark matter. We present a model for vector dark matter in an anisotropic Bianchi type I Universe.  Vector fields source anisotropies in the early universe characterized by a shear tensor which rapidly decays once the fields start oscillating, making them viable dark matter candidates. We present the set of equations needed to evolve scalar cosmological perturbations in the linear regime, both in synchronous gauge and Newtonian gauge.  We show that the shear tensor has to be taken into account in the calculation of adiabatic initial conditions. 
\end{abstract}

\maketitle

\section{Introduction}

Although the concordance model $\rm{\Lambda CDM}$ (which includes cold dark matter and a cosmological constant) does a great work at fitting cosmological observables, it does not explain the nature of the dark sector. Many alternative theories are being studied in order to explain the dark sector (DM), such as weakly interacting particles and ultralight fields (ULF). Recently, ULF models have gained interest due to its small scales predictions. Their small masses may let them behave as a collection  of waves, leading to new phenomena such as a suppression in the mass power spectrum on small scales, the presence of characteristic interference patterns,  the formation of soliton cores \cite{Marsh_2016,Ferreira_2021,Schive:2014dra}. One important characteristic of ULF models is that the cosmological observables are highly dependant on the particle mass.

In this work we focus on the study of cosmological perturbations in the linear regime. For the standard cold dark matter ($\rm{CDM}$) scenario, DM is treated as a pressureless component with no interactions other than gravity. We consider a vector field dark matter (VFDM) model, where DM is described by a classical real vector field with only gravitational interactions.  

One  important development  that motivates this study for VFDM models is the possibility of making very precise calculation to predict cosmological observables for the case of   scalar field dark matter (SFDM) candidates, which  serves  as the basis for the models at hand.   Indeed, for scalar ULF models, and not for vector ULF, there are public codes (such as AxionCamb \cite{Hlozek:2014lca} or class.FreeSF \cite{Ure_a_L_pez_2016}), which are based on the standard codes, CAMB\footnote{\url{http://camb.info}} or CLASS\footnote{\url{https://lesgourg.github.io/class_public/class.html}}. With the use of  such codes it is possible to calculate the evolution of the cosmological perturbations in the linear regime,  the power spectrum of the species, as well as many other cosmological observables that require the linear power spectrum, such as the temperature and polarization of the cosmic background radiation (CMB), the large-scale structure, the Lyman-alpha forest,  and hence to use precision cosmological data to set constrains on the models (see for instance \cite{Hlozek:2014lca,Ure_a_L_pez_2016,Pk_linear_lyman_1,Rogers:2020ltq,Kobayashi:2017jcf,Hlozek:2017zzf,Lague:2021frh,Dentler:2021zij,lin2023constraining,Lague:2023wes}). 
 
Specifically, the goal of this paper is to provide the complete set of equations together with an extension of the standard adiabatic initial condition that are necessary to   advance in the development of an extension of such codes, appropriate for the study of vector and mixed ULF models (that is, with different types of dark matter), which in turn will allow to contrast the predictions of theoretical models with recent observational data.

The other development motivating this study is that   non-scalar dark matter models are in principle as viable as the scalar ones. 
It has  been shown that there are several mechanisms that can give rise to the presence of a cosmological scalar ULF with large occupation numbers (for an introductory review see \cite{Marsh_2016}).  Although the production mechanisms for vector ULF are more restrictive (see for instance \cite{Arias:2012az,Graham_2016}),  recently several models have  been developed  \cite{Agrawal:2018vin,Dror:2018pdh,Co:2018lka,Long:2019lwl,Nakayama:2019rhg,Kaneta:2023lki,Kitajima:2023fun}.

The first difficulty in cases of spin fields already appears at the level of the background Universe, since such fields generally break isotropy. However, a general result \cite{Cembranos_2012_isotropy_theorem} shows that for massive fields without self-interactions and masses much larger than the expansion rate, the average energy-momentum tensor is isotropic and behaves like CDM\footnote{ Another possibility to preserve isotropy is to consider a triplet of orthogonal vector fields or N randomly oriented vector fields (see for example \cite{Golovnev_2008}). We will not consider these  alternatives in this paper.}.

In the early-universe, during radiation domination era one may expect the VFDM would not significantly affect the evolution of the metric and radiation perturbations. As we show in this work for adiabatic perturbations, although this expectation turns out to be  correct, to arrive at this conclusion it is necessary to consider an anisotropic background metric characterized by a tensor shear. Indeed, in  Sec. \ref{sec:adiabatic initial conditions} we show that the vector stress energy tensor has non-negligible contributions to the perturbed Einstein equations for superhorizon modes, at least for  masses $m\lesssim 10^{-22}\, \rm{eV}$.  Since the initial conditions for cosmological evolution of the perturbations are to be set when modes  are superhorizon and observables are in general sensitive to modifications of the initial conditions, if this contribution significantly modified  them, it  could leave an imprint on cosmological observables. However, it turns out that such VFDM contribution is exactly canceled by the presence of a term proportional to the shear.

The paper is organized as follows. In section \ref{sec:Vector field background} we present the VFDM model at the background level. We calculate the initial conditions and study the background evolution. In section \ref{sec:Vector field perturbations} we present the VFDM model at linear order in cosmological perturbation theory  in Synchronous gauge and we study the implications of the background shear in the calculation of adiabatic initial conditions.   
Finally, in section \ref{sec:Conclusions_results} we summarize the conclusions of our work.

In Appendix \ref{appendix:gauge_change} we present the metric in both gauges and discuss how the field and the metric perturbations transform  under a change of gauge. In Appendix \ref{apendix:synch_gauge} we present supplementary equations for the discussion of the Synchronous gauge, and in Appendix \ref{apendix:newton_gauge} we present the model in Newtonian gauge and the analogous  study of the initial conditions. In Appendix \ref{apendix:WeinbergAD} we use the  Weinberg's construction \cite{Weinberg_2003} to find the adiabatic solution for the vector field in the long wavelength limit. 

\section{Vector field background} \label{sec:Vector field background}

\subsection{Background equations}

We work in a Bianchi type I spacetime with metric given by

\begin{equation}
    ds^2 = a(\tau)^2 \left[ - d\tau^2 + \,\gamma_{ij} \, dx^i dx^j \right]\,,
\end{equation}

\noindent where $\gamma_{ij} = e^{-2\beta_i(\tau)} \delta_{ij}$ and the functions  $\beta_i$ (with $i=1,2,3$) are constrained by $\sum_i^3 \beta_i = 0$. Following \cite{Pereira_2007}, we define the shear of the metric as $\sigma_{ij} = \frac{1}{2}\gamma_{ij}^{\prime}$ where prime derivatives are with respect to conformal time $\tau$.

We consider a vector field dark matter (VFDM) model in the Bianchi I universe given by

\begin{equation}
    S = -\int d\tau\,dx^3\,\sqrt{-g}\left[\frac{1}{4} F^{\mu\nu}F_{\mu\nu} + \frac{m^2}{2} A^{\mu}A_{\mu}\right]\, ,
\end{equation}

\noindent where $F_{\mu\nu} = \nabla_{\mu}A_{\nu} - \nabla_{\nu}A_{\mu}$ is the usual field tensor, $g$ is the determinant of the metric and $m$ the vector mass. The vector field equations of motion are 

\begin{equation} 
    \nabla_{\nu} F^{\mu \nu} + m^2 A^{\mu} = 0\, .
    \label{general_eq_motion}
\end{equation}

In this work we consider that the VFDM candidate   can  be described as a combination of a background homogeneous field and a perturbation $A^{\mu}(\tau,\vec{x}) = A^{\mu}(\tau) + \delta A^{\mu}(\tau,\vec{x})$. In this section we focus on the background quantities. 

To solve for the homogeneous background component  we can split   Eq.~(\ref{general_eq_motion}) into an equation  for the spatial components $\vec{A}$, and a constraint equation for the time component giving $A^{0}=0$. The equations of motion are\footnote{ This equation was first derived in \cite{Ford_vector_inflation} in the context of inflation driven by a vector field.}

\begin{equation}
    A_{i}^{\prime\prime} - 2 {\sigma^k}_i\,A_k^{\prime} + m^2a^2 A_i = 0 \,.
    \label{eq_mov_A_background}
\end{equation}

The background fluid variables of the vector field in Bianchi I are
\begin{align}
    \rho_A &= \frac{1}{2 a^4} \left[ A_{i}^{\prime}A_{j}^{\prime} + m^2 a^2 A_{i}A_{j} \right]\,\gamma^{ij} \label{rho_A_background}\\[4pt]
    P_A &= \frac{1}{6 a^4} \left[ A_{i}^{\prime}A_{j}^{\prime} - m^2 a^2 A_{i}A_{j} \right] \,\gamma^{ij} \label{background_pressure_expression}\\[4pt]
    {\Sigma^i}_j &= \frac{1}{a^4} \bigg[ \frac{1}{3}A_{k}^{\prime} A_{l}^{\prime}\, \gamma^{kl} {\gamma^i}_j - A_{k}^{\prime} A_{j}^{\prime} \, \gamma^{ik}  \\
    &\qquad+ m^2 a^2 (A^i A_j - \frac{1}{3}A^k A_k {\gamma^i}_j)  \bigg]\,, \nonumber 
\end{align}

\noindent where ${\Sigma^i}_j$ is the traceless part of the energy-momentum tensor.  In what follows we use the convention of lowering and raising latin indices with the ``spatial metric" $\gamma_{ij}$ and its inverse $\gamma^{ij}$ respectively. The vector field is the only species  we consider here  that has anisotropies at the background level. Then, Einstein's nondiagonal equations are only sourced by the vector field,

\begin{equation}
    ({\sigma^i}_j)^{\prime} + 2 \mathcal{H}\, {\sigma^i}_j = \frac{a^2}{m_P^2} {\Sigma^i}_j\,,
    \label{metric_shear_eq}
\end{equation} 

\noindent with $m_P$ the Planck mass.

The last relevant equation for the background is Friedmann equation in Bianchi I geometry, given by

\begin{equation}\label{Friedmann}
    \mathcal{H}^2 - \frac{1}{6} \sigma^2 = \frac{a^2}{3 m_P^2} \rho\,,
\end{equation}

\noindent where $\sigma^2 = \sigma^{ij} \sigma_{ij}$. From the previous equation we define the shear relative abundance as

\begin{equation}
    \Omega_{\sigma} = \frac{\sigma^2}{6 \mathcal{H}^2} \,.
    \label{shear_abundance_def}
\end{equation}

\subsection{Initial conditions}\label{sec:initial_conditions_background}

The initial conditions for vector dark matter can be calculated by looking for  attractor solutions of the background equations   during radiation domination era where $a\propto \tau$. In what follows  we  assume  any initial anisotropy in the background metric can be neglected. We set the initial time $\tau_{\rm{ini}}$  early enough so that the mass of the vector field is sufficiently small with respect to the Hubble rate ($m\ll H_{\rm{ini}} \equiv \mathcal{H}_{\rm{ini}} a_{\rm{ini}}^{-1}$, where the subscript ``${\rm{ini}}$'' on a quantity denotes the quantity evaluated at the initial time). Otherwise, if $m\gtrsim H_{\rm{ini}}$,    it will be clear towards the end of this section that the   VFDM does not generate significant anisotropies on cosmological scales after the initial time.
 We solve the metric shear's equations (Eq.~\ref{metric_shear_eq}) perturbatively at linear order in the magnitude of  ${\Sigma^i}_j$   by keeping the lowest order in the metric shear on the right-hand side of Einstein's equations. That is, by setting the metric shear to zero on the fluid variables of the vector field. 
 
 In this work we assume an homogeneous  background vector  field  given by
\begin{equation}
    A_i(\tau) = A(\tau) \, \hat{A}_i \,.
\end{equation}

In Bianchi I the versors change in time because of time derivatives of $\gamma_{ij}$. In particular, we have that $\hat{A}_i^{\prime} \hat{A}^i = \sigma_A$, where $\sigma_A = \hat{A}^i \hat{A}^j \sigma_{ij}$. We solve the equations of motion of the vector at early times assuming that for any component of the shear we have that $|\sigma_{ij}| \ll \mathcal{H}$. Then, for the masses considered in this paper ($m\ll H_{\rm{ini}}$) we can solve Eq.~(\ref{eq_mov_A_background}) near the initial time as 

\begin{align}
    A_i(\tau) &\simeq (b + c \,\tau )\, \hat{A}_i\,,
    \label{cond_ini_vec_background}
\end{align}

\noindent where $b$ and $c$ are  integration constants. Therefore $A_i'(\tau)\simeq \mathcal{H}A_i(\tau)$ and   we can neglect the time derivatives of the versor. The constant mode corresponds to a decaying mode in the rescaled vector $\Bar{A}_i$ defined as $A_i = a\,\Bar{A}_i$. We set $b = 0$ for simplicity.  

The fluid variables at early times and in the lowest order approximation are given by

\begin{align}
    \rho_A &\simeq \frac{c^2}{2 a^4} \,,\label{rho_A_early_times}\\
    \Sigma_{ij} &\simeq - \frac{c^2}{a^4} \left(\hat{A}_i \hat{A}_j - \frac{\gamma_{ij}}{3} \right) \,,
\end{align}

\noindent with a radiationlike equation of state $w_A = \frac{1}{3}$.

Now we can solve for the shear.  Equation (\ref{metric_shear_eq}) admits the following formal solution

\begin{equation}
    {\sigma^i}_j = \left(\frac{a_{\rm ini}}{a}\right)^2 {S^i}_j + \frac{1}{a^2}\int_{a_{\rm{ini}}}^a s^3 \frac{{\Sigma^i}_j}{\mathcal{H} m_P^2} ds \,,
    \label{formal_solution_shear}
\end{equation}

\noindent  where ${S^i}_j$ is a constant tensor. We assume there is no significant shear at the initial time and set  ${S^i}_j=0$. By neglecting the metric shear in the vector stress-energy tensor, we obtain that the leading order solution can be written as (for  $a_{\rm{ini}}\ll a$)

\begin{equation}
    \sigma_{ij} \simeq - 6 R_A \mathcal{H} \left(\hat{A}_i \hat{A}_j - \frac{\gamma_{ij}}{3} \right) \,,
    \label{metric_shear_early_times}
\end{equation}

\noindent where  $R_A = \rho_A/\rho_{r}$ (with $\rho_{r}$ the radiation energy density) is the vector-to-radiation energy ratio. For the masses considered in this work at early times we have that $R_A \ll 1$. 

 Therefore, we obtained a shear that is much smaller than the Hubble rate, but which decays more slowly than any primordial shear that could eventually be present before $\tau_{\rm {ini}}$. As we show in Sec. IIIB, the magnitude of the shear generated at early times by the vector field is sufficiently large to have an impact in the  calculation of the initial condition for the cosmological perturbations for a range of masses $H_{eq} \lesssim m \lesssim 10^{-22} \rm{eV}$ (see Eq. 40), where $H_{eq}$   is the Hubble rate at equality.

\subsection{Background evolution}
 
In the previous section we have seen that the shear at early times has a decaying  mode determined by the vector-to-radiation energy ratio $R_A$. The general result of  \cite{Cembranos_2012_isotropy_theorem}  implies the VFDM effectively stops sourcing the metric shear after the Hubble rate falls below  the mass of the field. As a consequence the metric shear ${\sigma^i}_j$ decays as $a^{-2}$   at late times and the spacetime approaches to an isotropic  Friedman-Lemaître-Robertson-Walker (FLRW) one.
Indeed,  from Eq.~(\ref{eq_mov_A_background}) we can see that (at zero order in the shear) a few Hubble times after  $m a=m a_{osc}\sim \mathcal{H}$ the background  field  behaves as a an oscillatory function with a frequency given by the mass. Then, the background vector field has two different behaviors which are separated by the starting time for the  vector  oscillations (when the scale factor is $a=a_{osc}$). We can write the approximate  solution of  Eq.~(\ref{eq_mov_A_background})  as
\begin{equation}
    A(\tau) \propto 
    \begin{cases}
        a \qquad\qquad\qquad\qquad\,\,\,\,\,\, \quad a < a_{osc}\\
        a^{-\frac{1}{2}} \cos\left(\int ma \, d\tau\right) \,\quad\,  a_{osc} < a 
    \end{cases},
    \label{background_field_evolution}
\end{equation} where for $ a_{osc} < a $ we used a WKB approximation which assumes $m > H\equiv \mathcal{H}/a$. 

  In the regime  where the 
field oscillations become  fast on cosmological time-scales   the  effect of the fast oscillatory contributions in the VFDM stress-energy tensor can  be  computed by approximating the corresponding expression by  its time average.
We first notice that under the WKB approximation 
 we have that $\rho_A = (A^{\prime\,2}+ m^2a^2\,A^2)/2a^4 \propto a^{-3}$, at leading order in $H/m$.  Then,  the vector energy density at zero order in the shear can be approximated by  
\begin{equation}\label{rhoaback}
    \rho_A \simeq
    \begin{cases}
   \frac{ c^2}{2 a^4}  \quad\qquad\qquad\qquad \,\,a < a_{osc} \\
    \frac{ c^2}{2 a_{osc}^4}  \left(\frac{a_{osc}}{a}\right)^{3} \qquad a_{osc} < a
    \end{cases}.
\end{equation}
We see that once the field starts oscillating, it behaves as cold dark matter.  Indeed, by inserting Eq.~(\ref{background_field_evolution}) into (\ref{background_pressure_expression}) we can see that the pressure oscillates and averages to zero in this regime on cosmological timescales. The same argument holds for the shear tensor.

  By setting the initial conditions at $a_{\rm{ini}}\ll a_{osc}$, for $a_{\rm{ini}}\ll a<a_{osc}$ we can use  (\ref{metric_shear_early_times}) to   obtain  $\Omega_{\sigma}$ (given in  (\ref{shear_abundance_def}))  at leading order in $a_{\rm{ini}}/a$. Then, matching such solutions at $a=a_{osc}$  with the one obtained in the regime of fast oscillations of the VFDM (where the right-hand side of Eq.~(\ref{metric_shear_eq}) vanishes and  Eq.~(\ref{formal_solution_shear}) implies  $\sigma_{ij}$ decays as $a^{-2}$), we obtain:   
\begin{equation}
    \Omega_{\sigma} \simeq
    \begin{cases}
        4 \,R_A^2  \qquad\qquad\qquad  \,a < a_{osc} \\
        4 \,R_A^2 \left(\frac{H}{m}\right)^2 \,\,\qquad  a_{osc}  < a
    \end{cases}.
\end{equation} 
  
For the vector field to be the whole dark matter, it should  start oscillating before  matter-radiation  equality epoch \begin{equation}\label{boundaosc}
       a_{osc} < a_{eq}\simeq\frac{\Omega_{r,0}}{\Omega_{DM,0}} \,,    
\end{equation} where  $\Omega_{r}$ and $\Omega_{DM}$ are, respectively, the radiation and  the dark matter energy fraction, and the subscript $0$  denotes the  value at the present time. Using the Friedmann equation (Eq.~\ref{Friedmann}), for $a=a_{osc} < a_{eq}$   we can approximate\footnote{ For small values of $m$ a more precise calculation can be performed by including on the right-hand side  the vector contribution  and solving perturbatively for $a_{osc}$.}

\begin{equation}
    m^2 =H_{osc}^2    \simeq  H_0^2 \,\frac{\Omega_{r,0}}{ a_{osc}^{4}} \,,
\end{equation} 

\noindent with $H=\mathcal{H}/a$, from where we obtain

\begin{equation}\label{aosc}
    a_{osc}\simeq\Omega_{r,0}^{1/4}\left(\frac{H_0}{m}\right)^{1/2} \,.
\end{equation}

Using this result in Eq.~(\ref{rhoaback}) and that for $a>a_{osc}$,   $\rho_A=\rho_{DM}=3 m_p^2 H_0^2\Omega_{DM,0} a^{-3}$ we can solve for the constant $c^2$,\begin{equation}
    c^2\simeq 2\, \rho_{cr,0} \,\Omega_{DM,0}\, \Omega_{r,0}^{1/4}\left(\frac{H_0}{m}\right)^{1/2} \,,
\end{equation} where $\rho_{cr,0}= 3 m_p^2 H_0^2.$ 

 For  Planck cosmological parameters \cite{Planck:2018vyg},  from Eq.~(\ref{aosc}) and the condition $a_{osc} < a_{eq}$ we  
obtain $m > H_{eq} \sim 10^{-28} \rm{eV}$. Therefore, the mass range we consider in this paper is  $H_{ini} \gg m > H_{eq}$.  

At early times before the field starts oscillating ($a < a_{osc}$),  we obtain  $R_A = \rho_A/\rho_{r}\simeq  c^2/(2\rho_{r,0})$, and  $\Omega_{\sigma}$  can be written as \begin{equation}\label{shearconst}
    \Omega_{\sigma} \simeq 4 \, \Omega_{DM,0}^2 \Omega_{r,0}^{-3/2}\left(\frac{H_0}{m}\right)    \equiv \Omega_{\rm{ini}},
\end{equation}
\noindent where we have defined $\Omega_{\rm{ini}}$. For   $a > a_{osc}$ we have two different behaviors depending on whether we are on radiation or matter epoch,
\begin{equation}
    \Omega_{\sigma} \propto
    \begin{cases}
    \,\, a^{-2} \,\,\qquad a_{osc} < a < a_{eq}\, \\
    \,\, a^{-3} \,\,\,\qquad a_{eq} < a \, 
    \end{cases}  .
\end{equation}

One of the strongest constraints for the shear's abundance is set by \textit{big bang nucleosynthesis} (BBN), giving $\Omega_{\sigma} |_{BBN} \leq 10^{-3}$ for a universe with no anisotropic sources in the background \cite{Akarsu_2019}, where $\Omega_{\sigma}|_{BBN}$ is the shear abundance at BBN ($a \sim 10^{-8}$). Using this and Eq.~(\ref{shearconst}) it is immediate to obtain an approximate  lower bound on the VFDM masses allowed by BBN\footnote{Since for the  smaller masses allowed by this bound the vector starts oscillating closed to equality, in order to obtain a more precise bound one needs to include corrections in the background evolution, such as those mentioned in the previous footnote. A   precise background evolution  accounting for this corrections will be presented in \cite{paper_in_prep}.}:  
    
\begin{align}\label{LowerBBN}
    &m \gtrsim  4 \times 10^3 \, \Omega_{DM,0}^2 \, \Omega_{r,0}^{-3/2} H_0 \,\\[4pt]
    &\sim 0.25\times 10^{-24} \rm{eV} \left( \frac{\Omega_{DM,0}}{0.26} \right)^2 \left( \frac{\Omega_{r,0}}{10^{-4}} \right)^{-\frac{3}{2}} \left( \frac{H_0}{10^{-33} \rm{eV}} \right) \,,\nonumber
    \end{align}
 where to write the second line we used approximate values $\Omega_{DM,0}\sim 0.26$, $\Omega_{r,0}\sim{10^{-4}}$, ${H_0}\sim {10^{-33} \rm{eV}} $, estimated from Planck   cosmological parameters   \cite{Planck:2018vyg}. For such parameters    we obtain the bound  $m \gtrsim    0.25\times 10^{-24} \rm{eV}$.  
 
   By analogy with the scalar field case, stronger bounds on $m$ are expected from the analysis of   cosmological perturbations, even in the linear regime.  Interestingly, as we show next, a simple estimate indicates  that for small values of the mass (but still allowed by the BBN constraint in Eq.~(\ref{LowerBBN})) the characteristic anisotropy of the VFDM case (which is absent in the scalar case) may have an impact on the CMB   quadrupole temperature.  Notice  VFDM models are different from the standard CDM scenario in a Bianchi I background metric considered in   \cite{Akarsu_2019}. For  the latter scenario the authors of \cite{Akarsu_2019} conclude that   given the BBN constraint on the initial $\Omega_{\sigma}$,   the effect on the  CMB quadrupole temperature turns out to be observationally irrelevant. In particular, they estimated a value of $\Omega_{\sigma}$ at the present time,  given by $\Omega_{\sigma,0}\sim 4 \times 10^{-20}\equiv \Omega_{\sigma,q}$, for which the   so called quadrupole temperature problem \cite{Planck:2018vyg} could be addressed. For smaller values of $\Omega_{\sigma,0}$ the effect on the  quadrupole would be smaller. The effect of the metric shear on the quadrupole temperature  depends on the evolution of the quantity $\Omega_{\sigma}$ from the time of decoupling of CMB photons to the present (see  \cite{Akarsu_2019}). Since during such period of time  the VFDM evolves effectively as CDM (at the background level), the background metric shear has no source and the quantity $\Omega_{\sigma}$  evolves just as in the standard Bianchi I scenario studied in \cite{Akarsu_2019}.
The reason why in the VFDM scenario  such conclusion does not apply for all of the masses of the VFDM  we considered here goes as follows. For  masses smaller than the Hubble  rate at BBN ($m<H_{BBN}$)\footnote{For $m>H_{BBN}$, the field effectively behaves as CDM at the background level and, therefore, the same argument given in \cite{Akarsu_2019} does apply.},  in the VFDM scenario the  anisotropy characterized by  $\Omega_{\sigma}$ decays   less from BBN to the present  than in the standard case. While in the standard case $\Omega_{\sigma}$ always decays (as $a^{-2}$ in radiation domination era and as $a^{-3}$ in matter domination era), in the VFDM scenario   $\Omega_{\sigma}$ is practically  constant from BBN until the field starts oscillating   (i.e., when $H\simeq m$). Indeed, by matching the different approximations for the evolution of  $\Omega_{\sigma}$, in the VFDM scenario, we  have
\begin{equation}
    \Omega_{\sigma} \simeq
    \begin{cases}\label{omegasigmaest}
      \,\,   \Omega_{\rm{ini}}\left(\frac{a_{osc}}{a}\right)^2  \,\,\qquad a_{osc} < a < a_{eq}\, \\[4pt]
     \,\,   \Omega_{\rm{ini}}\left(\frac{a_{osc}}{a_{eq}}\right)^2 \left(\frac{a_{eq}}{a}\right)^3  \,\,\,\qquad a_{eq} < a \,  \\ 
    \end{cases} ,
\end{equation} 
 where  $\Omega_{\rm{ini}}$, $a_{eq}$ and $a_{osc}$ are  given in  (\ref{shearconst}),    (\ref{aosc}) and  (\ref{boundaosc}), respectively.  Hence we can estimate  $\Omega_{\sigma,0}$ in the VFDM model (ignoring the effect of the cosmological constant) by setting $a=1$ on the second line of Eq.~(\ref{omegasigmaest}). Then, the condition   $\Omega_{\sigma,0}\gtrsim  \Omega_{\sigma,q}$ under which we     expect a relevant  effect on the CMB quadrupole temperature can be recast as the following  bound in the mass of the VFDM:
\begin{align}\label{CMBquqd}
  m &\lesssim 2 H_0  \sqrt{\frac{\Omega_{DM}}{\Omega_{\sigma, q}}}\\
  &\sim 5 \times 10^{-24} \rm{eV}
  \left(\frac{H_0}{10^{-33} \rm{eV}}\right)  \sqrt{\frac{\Omega_{DM}}{0,26}}   \sqrt{\frac{4 \times 10^{-20}}{\Omega_{\sigma, q}}}.\nonumber
\end{align}
Therefore, there is a mass range for which the BBN constraint as estimated in (\ref{LowerBBN}) can be satisfied and the VFDM can induce  a relevant effect on the CMB quadrupole temperature.  We expect to quantify this more precisely in future work. For this, the equations presented below are the basic necessary ingredients.

\section{Vector field perturbations} \label{sec:Vector field perturbations}

In this section we focus on Einstein's equations at linear order in perturbation theory.  We show that the metric shear has to be considered in the calculation of adiabatic initial conditions.  We work here in synchronous gauge and  we present the equations in Newtonian gauge in Appendix \ref{apendix:newton_gauge}. As it will be enough, in what follows we only keep up to linear corrections in the shear and assume that $|\sigma_{ij}| \ll \mathcal{H}$.

We work with the perturbations in Fourier space, and use the convention

\begin{equation}
    f(\vec{x}) = \int d^3k\,f(\vec{k})\,e^{i\,\vec{k}\cdot\vec{x}}\,,
\end{equation}

\noindent   where   we denote with ``$\cdot$'' the product with metric $\gamma_{ij}$ (e.g. $\vec{k}\cdot\vec{x}=k_ix^i$). Here $x^i$ are the comoving coordinates. In Fourier space $k_i$ is constant while $k^i\equiv\gamma^{ij}k_i$ changes with time. 

It is convenient to choose an orthonormal basis given by $\{ \hat{e}_1, \hat{e}_2, \hat{e}_3 \}$ where $\hat{e}_3 = \hat{k}$, $\hat{e}_2 = \hat{k} \times \hat{A}$ and $\hat{e}_1 = \hat{k} \times \hat{e}_2$.  This is a mode-dependent basis where the background vector is always contained in the plane defined by $\hat{k}$ and $\hat{e}_1$. We decompose the metric shear in this basis as

\begin{equation}
    \sigma_{ij} = \frac{3}{2} \left( \hat{k}_i \hat{k}_j - \frac{\gamma_{ij}}{3} \right) \sigma_{\parallel} + 2\,\sum_{a=1,2} \sigma_{v_a} \, \hat{k}_{(i}\,\hat{e}_{j)}^a + \sum_{\lambda=+,\times} \sigma_{\lambda} \,\epsilon_{ij}^{\lambda} \,,
\end{equation}

\noindent where $\epsilon_{ij}^{+} = \hat{e}_i^1 \,\hat{e}_j^1 - \hat{e}_i^2 \,\hat{e}_j^2$ and $\epsilon_{ij}^{\times} = \hat{e}_i^1\, \hat{e}_j^2 + \hat{e}_i^2 \,\hat{e}_j^1$. By comparing with Eq.~(\ref{metric_shear_early_times}) it can be shown that in this basis, the shear at early times is given by

\begin{subequations}
\begin{align}
    \sigma_{\parallel} &= \frac{-2 c_L^2 + c_{t_1}^2}{3 m_P^2 a^2 \mathcal{H}}\,, \label{sigmaparalel}\\[3pt]
    \sigma_{v_1} &= \frac{c_L c_{t_1}}{m_P^2 a^2 \mathcal{H}}\,, \\[3pt]
    \sigma_{+} &= - \frac{c_{t_1}^2}{2 m_P^2 a^2 \mathcal{H}}\,,
\end{align}
\end{subequations}

\noindent where $c_L = c\, (\hat{k}\cdot \hat{A})$ and $c_{t_1} = c\, (\hat{e}_1\cdot \hat{A})$. The remaining components of the shear are not sourced by the vector at early times, so we can set $\sigma_{v_2} = \sigma_{\times} = 0$.

Taking into account that $\gamma^{ij}$ is time dependent  the derivative of the relevant versors can be derived as\footnote{ For a more detailed derivation see Sec. II in \cite{Pereira_2007}}. 

\begin{subequations}
\begin{align}
    k^{\prime}\,\,\,\, &= - \sigma_{\parallel} \, k\,, \\[6pt]
    (\hat{k}_i)^{\,\prime} &= \sigma_{\parallel} \, \hat{k}_i\,, \\[6pt]
    (\hat{e}^a_i)^{\,\prime} &= - \sum_b \sigma_{lj} e_a^l e_b^j \, e^b_i + 2 \sigma_{ij} e^j_a \,.
\end{align}
\end{subequations}

\subsection{Linear equations}   \label{sec:Vector field perturbations synch}

Now we present Einstein's equations in synchronous gauge as defined in Appendix \ref{appendix:gauge_change}, which in Fourier space reads

\begin{subequations}
\begin{align}
   \delta g_{00} &= \delta g_{0i} = 0\,,\\[7pt]
   \delta g_{ij} &= a^2\left[-2\left(\gamma_{ij}+\frac{\sigma_{ij}}{\mathcal{H}}\right)\, \eta + \hat{k}_i\hat{k}_j (h+6\eta) \right]\,,
\end{align}    
\end{subequations}

\noindent where we have neglected vector and tensor perturbations of the metric. For simplicity, we consider the equations to linear order in any component of $\sigma_{ij}$.

The vector equations of motion at linear order can be extracted from Eq.~(\ref{general_eq_motion}).  In order to do this we wrote a code in mathematica   using xAct - xPand packages \cite{Pitrou:2013hga,xact}. As for the background, the dynamics can be split into a constraint equation for the temporal component and one equation of motion for each polarization of the spatial components. By projecting on the basis we obtain 

\begin{subequations} 
\begin{align}
    &\delta A_L^{\prime\prime} + (m^2 a^2 + \sigma_{\parallel}^{\prime}) \delta A_L - i\,k \left(\delta A_0^{\prime} - 2 \sigma_{\parallel} \delta A_0\right) = S_L\,,
    \label{eq_mov_delta_A_L_synch}\\
    &\delta A_{t_1}^{\prime\prime} + (m^2 a^2 + k^2  - \frac{\sigma_{\parallel}^{\prime}}{2} + \sigma_{+}^{\prime} ) \delta A_{t_1} + 2i\,k\, \sigma_{v_1} \delta A_0 = S_{t_1}\,,\
    \label{eq_mov_delta_A_T_1_synch}\\
    &\delta A_{t_2}^{\prime\prime} + (m^2 a^2 + k^2  - \frac{\sigma_{\parallel}^{\prime}}{2} -  \sigma_{+}^{\prime} ) \delta A_{t_2} = 0\,,
    \label{eq_mov_delta_A_T_2_synch}
\end{align}
\label{eq_mov_delta_A_synch}
\end{subequations}

\noindent where the sources on the right-hand side are given by Eq.~(\ref{sources_synch}). From the temporal equation we have that

\begin{align}
    \delta A_0 &= \frac{-i\,k}{m^2 a^2 + k^2} \, \big[\delta A_L^{\prime} + \sigma_{\parallel} \delta A_L + 2 \sigma_{v_1} \delta A_{t_1} \\[5pt]
    &+ 2 \frac{\sigma_{v_1}}{\mathcal{H}} A_{t_1}^{\prime} \eta - \frac{1}{2}(A_L^{\prime} + \sigma_{\parallel} A_L + 2 \sigma_{v_1} A_{t_1}) (h + 8 \eta ) \big]\,.   \nonumber
\end{align}

The Einstein's equations to linear order are

\begin{subequations} 
    \begin{align}
        &k^2 \eta - \frac{1}{2} \mathcal{H} h^{\prime} + \frac{3}{2} \sigma_{\parallel} \eta^{\prime} = -\frac{a^2}{2 m_P^2} \delta \rho \label{einstein_00_synch}\,,\\[7pt] 
        &k^2 \eta^{\prime} + \frac{k^2}{2} \left[\left(\frac{\mathcal{H}^{\prime}}{\mathcal{H}^2} - 3\right) \sigma_{\parallel} - \frac{\sigma^{\prime}_{\parallel}}{\mathcal{H}} \right] \eta = \frac{a^2}{2 m_P^2} (\rho + P) \theta \label{einstein_0i_synch}\,,\\[7pt]
        & h^{\prime\prime} + 2 \mathcal{H} h^{\prime} - 2 k^2 \eta + 9\, \sigma_{\parallel} \eta^{\prime}  = -\frac{3\, a^2}{m_P^2} \delta P\,, \label{einstein_ii_synch} \\[7pt]
        &h^{\prime\prime} + 6 \eta^{\prime\prime} + 2 \mathcal{H} (h^{\prime} + 6 \eta^{\prime}) - 2 k^2\, \eta +\nonumber 3\bigg[ \frac{\sigma_{\parallel}^{\prime\prime}}{\mathcal{H}} - \frac{\mathcal{H}^{\prime\prime}}{\mathcal{H}^2} \sigma_{\parallel}  \\
        &- 2(\frac{\mathcal{H}^{\prime}}{\mathcal{H}^2}-1)(\sigma_{\parallel}^{\prime} - \frac{\mathcal{H}^{\prime}}{\mathcal{H}}\sigma_{\parallel}) \bigg]\eta = -\frac{3 a^2}{m_P^2}\, (\rho + P) \delta\Sigma_{\parallel}\,, \label{einstein_ij_synch}
    \end{align}
    \label{einstein_eqs_synch}
\end{subequations}

\noindent where the fluid variables are defined as

\begin{subequations}
\begin{align}
    \delta\rho &= - \delta {T^0}_0 \,,\\[6pt]
    \delta P &= \frac{1}{3}\delta {T^i}_i\,, \\[6pt]
    (\rho + P) \theta &= i k^i \delta {T^0}_i \,,\\[6pt]
    (\rho + P) \delta\Sigma_{\parallel} &= - \big(\hat{k}_i\hat{k}^j - \frac{1}{3} {\gamma^j}_i\big) \delta {\Sigma^i}_j\,,
\end{align}
\label{fluid_variables_definitions}
\end{subequations} 

\noindent and the expressions in terms of the vector field are given in Appendix \ref{apendix:synch_gauge}. 

By setting $\sigma_{\parallel} = 0$ in Eqs.~(\ref{einstein_eqs_synch}) we recover the FLRW limit (see for instance \cite{Ma_1995}).

\subsection{Adiabatic initial conditions} \label{sec:adiabatic initial conditions}

In the \rm{$\Lambda$}CDM cosmological model the initial conditions for Einstein's equations are calculated by extending the solutions far outside the horizon ($k\tau \ll 1$) and deep in the radiation era. In this regime the radiation dominates the right-hand side of Einstein's equations, so the rest of the species can be neglected. The equations are then solved by looking for the attractor solutions of the radiation and the metric. These solutions can be found from Eqs.~(\ref{einstein_00_synch}), (\ref{einstein_ii_synch}) and (\ref{einstein_ij_synch}) in this regime, and are given by

\begin{subequations}\label{adiabinicond}
\begin{align}
    \eta &= \eta_0 - \alpha_R \frac{\eta_0}{2} (k \tau)^2\,,\label{ecetaLCDM}\\
    h &= \frac{\eta_0}{2} (k \tau)^2\,, \label{h_ini} \\
    \delta_{\gamma} &= - \frac{\eta_0}{3} (k\tau)^2 \,,
\end{align}
\label{h_eta_lcdm_cond_ini}
\end{subequations}

\noindent where $\delta_{\gamma}=\delta\rho_{\gamma}/\rho_{\gamma}$ (with the subscript  $\gamma$ denoting  photons),  $\alpha$ is a constant that is fixed with Eq.~(\ref{einstein_0i_synch}) (in $\Lambda$CDM  $\alpha_R = (5 + 4 R_{\nu})(15 + 4 R_{\nu})^{-1}/12$ with $R_{\nu}$ the ratio of neutrinos to radiation energy density   \cite{Ma_1995}) and $\eta_0$ is an integration constant. The initial conditions for the rest of the species are calculated by imposing adiabatic initial conditions $\frac{\delta \rho_i}{ \dot{\rho}_i} = \frac{\delta \rho_j}{ \dot{\rho}_j}$, where the index runs over every species.

In the presence of a background anisotropic stress the metric shear $\sigma_{\parallel}$ has to be included in this calculation. At early times we have that $\sigma_{\parallel} \ll \mathcal{H}$, so the shear can be neglected from the left-hand side of Eqs.~(\ref{einstein_00_synch}), (\ref{einstein_ii_synch}) and (\ref{einstein_ij_synch}). 
In this section we show that in the case of adiabatic initial conditions, we can also neglect the vector contribution from the respective equations.
Then, we can solve for the metric perturbations $h$ and $\eta$ and the radiation energy density as in  $\rm{\Lambda CDM}$ model. 
However, in the $k \ll \mathcal{H}$ regime  the leading order of Eq.~(\ref{einstein_0i_synch}) in   powers of $k$ is dominated by the shear in the left-hand side (as can be easily seen by inserting  Eq.~(\ref{ecetaLCDM}) into  Eq.~(\ref{einstein_0i_synch})), giving a contribution scaling as $k^2$.  Indeed, working at  leading order in $k\tau$ for superhorizon modes and using Eq.~(\ref{metric_shear_early_times}),   
the  term with the shear becomes subdominant only when 

\begin{align} \label{EqmodesIR}
    (k\tau)^2 &\gtrsim  9 \,\alpha_R^{-1} R_A\left| \cos^2\theta_k-\frac{1}{3}\right|\\
    &\sim 9  \frac{\Omega_{DM,0}}{\Omega_{r,0}^{3/4}\alpha_R }\left(\frac{H_0}{m}\right)^{1/2}\left| \cos^2\theta_k-\frac{1}{3}\right|, \nonumber
\end{align} 

\noindent where $\cos\theta_k=\hat{A}\cdot \hat{k}$.   
In order to quantify the effect of the shear on observable quantities such as CMB one can study the evolution of the perturbations for the same initial conditions with and without the shear. This can be done in an  extension of a standard  Einstein-Boltzmann code which implements the equations presented in this paper. We are currently working on such extension \cite{paper_in_prep} and on this study. For the moment  we notice that if one sets the initial condition outside horizon at most at $k\tau=k\tau_i\simeq 0.1$,  Eq.~(\ref{EqmodesIR}) implies that the  shear can   be neglected for masses  such that
\begin{align} 
    m  &\gtrsim   0.55 \times 10^{-22} \rm{eV} \left(\frac{0.1}{ k\tau_i}\right)^4  \frac{\left(\frac{\Omega_{DM,0}}{0.26}\right)^{2}}{ \alpha_R^2 (10^4 \,\Omega_{r,0})^{3/2}}\, \nonumber\\
    &\quad\times\left(\frac{H_0}{10^{-33} \rm{eV}}\right)\left| \cos^2\theta_k-\frac{1}{3}\right|^{2}. 
    \label{range_of_masses_ad_ic}
\end{align} 

 Hence,  for  $\Omega_{DM,0}\sim 0.26$, $\Omega_{r,0}\sim{10^{-4}}$, ${H_0}\sim {10^{-33} \rm{eV}} $,   we expect the effect to be potentially significant   for  $m\lesssim 10^{-22} \rm{eV}$.
 
Let us now analyze the right-hand side of Eq.~(\ref{einstein_0i_synch}). 
The initial condition  for the velocity gradients,  $\theta_i$, can be  obtained from the  fluid-like equations   for each species  in the long wavelength regime as described in \cite{Ma_1995}. For instance, for photons, from the equations  at leading order in $k \tau$, 

\begin{subequations}
\begin{align}
&\delta_\gamma'+\frac{4}{3}\theta_\gamma+\frac{2}{3}h'=0\,,\\
&\theta_\gamma'-\frac{k^2}{4}\delta_\gamma=0\,,
\end{align}
\end{subequations}

we obtain   

\begin{equation}\label{velgradphotons}
    \theta_\gamma = -\frac{\eta_0}{36} k^4 \tau^3\,.
\end{equation} 

Using  this result and Eq.~(\ref{adiabinicond}) into Eq.~(\ref{einstein_0i_synch}), from the  scaling behavior in powers of $k^2$ of the terms it is immediate to conclude that  to preserve the standard adiabatic initial conditions for the metric and radiation perturbations  also in the infrared (when the contribution of the shear is important), the vector has to cancel the shear's contribution at early times.

The attractor solutions for the vector are found by solving the equation of motion at early times far outside the horizon. The solutions for Eqs.~(\ref{eq_mov_delta_A_L_synch}), (\ref{eq_mov_delta_A_T_1_synch}) and (\ref{eq_mov_delta_A_T_2_synch}) in this regime and for the masses considered in this work can be written as a power law of the conformal time. The   integration constants can be calculated by imposing adiabatic initial conditions with the radiation, namely  $\delta_A = \frac{3}{4} \delta_{\gamma} (1 + w_A)$ at early times. At zero order in the metric shear we obtain

\begin{equation}
    \delta \vec{A} =   \left(2 c_L\, \hat{k} - c_{t_1}\,\hat{e}_1 \right) \tau\,\eta_0\,.
\end{equation}

With this initial condition we can now check that the scaling of the leading order in $k\tau$ of $\delta_A$, $\delta P_A$ and $\delta \Sigma_{A\parallel}$ (from Eqs.~(\ref{delta_rho_synch}), (\ref{delta_P_synch}) and (\ref{delta_shear_synch}), respectively) is the same as that of the other species (that is, $\sim k^2\tau^2$). However, this contribution is suppressed by a factor of $R_A = \frac{\rho_A}{\rho_{r}}$ in Einstein's equations (Eqs. (\ref{einstein_00_synch}), (\ref{einstein_ii_synch}) and (\ref{einstein_ij_synch}), respectively). Then, we can neglect the vector  variables from these equations at early times.  
The initial condition for $\theta_A$ is given by 

\begin{equation} 
    (\rho_A+P_A)\theta_A = \frac{2 c_L^2 - c_{t_1}^2}{a^4}\,k^2\tau\,\eta_0\,,
    \label{theta_A_ini_synch} 
\end{equation}
 
\noindent and therefore, inserting this result into Eq.~(\ref{einstein_0i_synch}) and using  (\ref{sigmaparalel}), we see that this vector contribution exactly cancels the one of the metric shear at early times.

The dark matter is expected to be subdominant at early times, but comparing Eqs.~(\ref{velgradphotons}) and (\ref{theta_A_ini_synch}) far outside the horizon we see that the vector  velocity divergence dominates over the one of radiation in this regime. However, in the presence of vector dark matter it is necessary to consider an anisotropic metric, and the contribution of such large vector velocity divergence    is exactly canceled by the  contribution of the shear, so  that the  standard adiabatic initial conditions are  recovered at early times. Otherwise, if one insists on working in an isotropic universe,  in the calculation of the initial conditions for the radiation and the metric
 one would erroneously find a large infrared effect  on the   perturbations  due to the   large   velocity gradient $\theta_A$ of the vector field. In Newtonian gauge it can be shown (see Appendix \ref{apendix:newton_gauge}) that the vector shear $\delta \Sigma_A$ dominates Einstein's traceless-longitudinal equation. As in the synchronous gauge, it is necessary to consider an anisotropic metric to cancel this contribution.

\section{Conclusions}\label{sec:Conclusions_results}

In this work we have computed all equations needed to study the evolution of scalar cosmological perturbations in the linear regime in VFDM models, neglecting vector and tensor perturbations. The consistency of the model requires the background metric must be taken to be an anisotropic  Bianchi I metric, characterized by a metric shear produced by a VFDM background pointing in a fixed direction. We have presented the equations for the metric and VFDM perturbations around such anisotropic background, in both  synchronous gauge (see Eqs.~(\ref{einstein_eqs_synch}) and ~(\ref{eq_mov_delta_A_synch}) with the sources given in  Eq.~(\ref{sources_synch}))  and Newtonian gauge (see Eqs.~(\ref{einstein_eqs_new}) and ~(\ref{eq_mov_delta_A_synch}) with the sources given in  Eq.~(\ref{sources_newt})).

We have studied the perturbations in the long-wavelength (superhorizon) regime  at early times during radiation domination. We have shown that   the limit  $k\to0$ with $k\neq0$ is subtle. Indeed, for small enough values of $k$,  the contribution of the metric  shear becomes of the same order as the other terms in the left-hand side of Einstein equations. In the same limit   the  right-hand side of Einstein equations involves large VFDM contribution (in synchronous gauge this corresponds to a VFDM velocity gradient  that becomes larger than the velocity gradient of photons, while in the Newtonian gauge the anisotropic VFDM shear does not vanish as happens for the standard species). Therefore, a careful study of  the evolution of both the metric shear and the VFDM in that regime,  as the one presented in this paper,  becomes crucial to appropriately set  the initial conditions for cosmological perturbations. We have derived the adiabatic initial conditions for the VFDM, and we have shown the relevant contributions of the metric shear and the one of the VFDM cancel each other, so that the initial condition for the metric perturbations   and the perturbations of the other species remain the same as in \rm{$\Lambda$}CDM  at leading order in the  the vector-to-radiation energy ratio $R_A$. 

Our analytic study of the evolution indicates VFDM models are in principle as viable as the SFDM one. An important difference that is already present at the background level is the contribution of the metric shear to the Friedman equation, given by the corresponding abundance $\Omega_{\sigma}$. In this paper we  have used  known constrains   on $\Omega_{\sigma}$ from  BBN \cite{Akarsu_2019} to estimate a lower bound on the VFDM mass $m$ (see Eq. \ref{LowerBBN}).  We have  also used the results of \cite{Akarsu_2019}  to estimate the mass range for which one can expect the VFDM to affect the CMB temperature quadrupole in an observationally significant way, without affecting BBN predictions. The estimated  mass range is between the BBN bound given in (\ref{LowerBBN}), and the masses satisfying Eq.~(\ref{CMBquqd}). Moreover, in view of our knowledge of the SFDM case (see for instance \cite{Hlozek:2014lca,Hlozek:2017zzf}), in the VFDM case we expect to obtain stronger bounds from a quantitative study of the evolution of the cosmological perturbations, even in the linear regime.  The equations derived in this paper are a  necessary ingredient to start  such study precisely. The following step to make progress in this direction is to numerically implement the equations in a code  such as CAMB or CLASS, which is the main goal  of \cite{paper_in_prep}. 
Finally, in the future, it would be worth to extend the equations to include vector and tensor perturbations which  mix themselves in the presence of the VFDM anisotropic background. 

\begin{acknowledgments} 
   We would like to thank Juan Manuel Armaleo, Susana Landau and Matias Leizerovich for discussions.   
This work has been supported by CONICET and UBA. We acknowledge the use of the xAct - xPand package for Mathematica \cite{Pitrou:2013hga,xact} 
\end{acknowledgments}

\appendix

\section{Gauge change in anisotropic background} \label{appendix:gauge_change}
 In this Appendix we study change of coordinates in Bianchi I geometries. In particular, we consider the change of gauge between Newtonian gauge and synchronous gauge. The most general metric perturbation in Bianchi I is given by
\begin{equation}\label{genericgauge}
    ds^2 = -a^2(1+2\phi) d\tau^2 + a^2 2B_i \, dx^i d\tau + a^2\left( \gamma_{ij}  + h_{ij}\right) dx^i dx^j
\end{equation}
\noindent where we can decompose
\begin{align}
    B_i &= \partial_i B + V_i\,, \\
    h_{ij} &= - 2 \left(\gamma_{ij} + \frac{\sigma_{ij}}{\mathcal{H}}\right)\psi + 2\,\partial_i \partial_j E + 2\, \partial_{(i} E_{j)} + 2\,{E}_{ij}\,, \label{metric_pert_h_ij}
\end{align}
\noindent with $\partial_i V^i = \partial_i E^i = 0$ and ${{E}^i}_i = \partial_i {{E}^i}_j = 0$. We consider a change of coordinates 
\begin{equation}
    \hat{x}^{\mu} = x^{\mu} + \xi^{\mu}\,,
\end{equation}
\noindent where $\xi^{\mu}$ is order one in perturbation theory and can be written as $\xi^0 \equiv \alpha$ and $\xi^i \equiv \beta^{,i} + \epsilon^i$ with ${\epsilon^i}_{,i} = 0$. The metric transforms under this change of coordinates as
\begin{equation}
    \hat{g}_{\mu\nu}(x) = g_{\mu\nu}(x) - g_{\mu\sigma}(x)\,\partial_{\nu}\xi^{\sigma} - g_{\nu\sigma}(x)\,\partial_{\mu}\xi^{\sigma} - \xi^{\sigma}\partial_{\sigma} g_{\mu\nu}(x)\,.
    \label{gauge_change_metric_def}
\end{equation}

The transformation of the metric perturbations defined in (\ref{genericgauge}) can be found by inspection of the components of the previous equation. The temporal equation, the $0i$ equation respectively, the trace of that with   spatial indices and  the projection on $(\hat{k}_i \hat{k}_j - \frac{1}{3}\gamma_{ij})$ directions,  yielding to
\begin{subequations}
\begin{align}
    \hat{\phi} &=  \phi - \alpha^{\prime} - \mathcal{H}\,\alpha\,, \\[4pt]
    \hat{\psi}&= \psi + \mathcal{H}\,\alpha\,,  \\[4pt]
    \hat{E} &= E - \beta \,,\\[4pt]
    \hat{B} &= B + \alpha - k^{-2} \left( k^2 \beta\right)^{\prime}\,. 
\end{align}
\end{subequations}
 
The transformation properties of the vector perturbations can be calculated by multiplying  Eq.~(\ref{gauge_change_metric_def}) by $P_{ij} = (\gamma_{ij} - \hat{k}_i\hat{k}_j)$, which selects the component in a direction perpendicular to $\hat{k}$. Then, it can be shown that
\begin{align}
    \hat{V}_i &= V_i - \gamma_{ij} (\epsilon^j)^{\prime} + 2 i \sigma_{lj} k^j {P^l}_i \,\beta \\[4pt]
    \hat{E}_i &= E_i - \epsilon_i \,.
\end{align} 

Finally, the tensor perturbation is gauge invariant, so $\hat{E}_{ij} = E_{ij}$. This is a consequence of introducing the shear in the definition of the metric perturbation $h_{ij}$ in the first term of Eq.~(\ref{metric_pert_h_ij}).

The functions $\alpha$ and $\beta$ can be found by fixing the gauge. In this work we are interested in transformations between synchronous and Newtonian gauge. We define the $\hat{x}^{\mu}$ coordinates to be in synchronous gauge while $x^{\mu}$ coordinates are in Newtonian gauge. The Newtonian gauge is given by the metric potentials $\phi$ and $\psi$ and by setting $E = E_i = 0$. For the synchronous gauge, we use the convention of \cite{Ma_1995} and define it by fixing $\hat{\phi} = \hat{B}_i = 0$ and by defining two new scalar functions $h$ and $\eta$ so that the spatial part of the metric is written as\footnote{Einstein-Boltzmann solver codes are generally written in terms of this scalar functions $h$ and $\eta$.}
\begin{align}
    \hat{\psi} &= \eta\,, \\[4pt]
    \hat{E} &= -\frac{1}{2k^2} (h + 6 \eta)\,.
\end{align}
 In terms of these functions $h$ and $\eta$,  the change of coordinates is given by  

\begin{align}
    \alpha &= \beta^{\prime} - 2 \sigma_{\parallel}\, \beta\,, \\[4pt]
    \beta &= \frac{1}{2k^2} (h + 6 \eta)\,, \\[4pt]
    \epsilon_i &= - E_i\,.
\end{align}

The energy momentum tensor transforms as a rank-2 tensor, so it has the same transformation properties as the metric (Eq.~\ref{gauge_change_metric_def}). By inspection of the different components we obtain   
\begin{subequations}
    \begin{align}
        \Delta {T^0}_0 &= - \rho^{\prime} \,\xi^0\,, \\[4pt]
        \Delta {T^0}_i &= (\rho + P) {\xi^0}_{,i} + {\Sigma^k}_i {\xi^0}_{, k}\,. \\[4pt]
        \Delta {T^i}_j &= ({\xi^i}_{,k} {\Sigma^k}_j - {\xi^k}_{,j} {\Sigma^i}_k) - (P^{\prime} {\delta^i}_j + {{\Sigma^{\prime}}^i}_j)\, \xi^0 \,,
    \end{align}
\end{subequations}
It is now straightforward to show that the fluid variables thus changes as  
\begin{subequations}
    \begin{align}
        \delta \hat{\rho} &= \delta \rho - \rho^{\prime}\, \alpha \,,\\[4pt]
        \delta \hat{P} &=  \delta P - P^{\prime} \,\alpha \,,\\[4pt]
        (\rho+P)\hat{\theta} &= (\rho+P)\theta -  k^2 \left[(\rho + P) + \hat{k}_l \hat{k}^i {\Sigma^l}_i\right] \alpha \,,\\[4pt]
        \delta \hat{\Sigma}_{\parallel} &=   \delta \Sigma_{\parallel} + i k_i \epsilon^j {\Sigma^i}_j - (\hat{k}_i \hat{k}^j + \frac{1}{3}{\gamma^j}_i){{\Sigma^{\prime}}^i}_j\,\alpha\,. 
    \end{align}
\end{subequations} 
Finally, the vector field transforms under a change of coordinates as
\begin{equation}
    \delta{\hat{A}}_{\mu}({\hat{x}}) = \delta A_{\mu}(x) -  \partial_{\mu} {\xi}^{\nu}  A_{\nu}(x) -  \partial_{\nu} A_{\mu}(x) \xi^{\nu}\,,
\end{equation}
\noindent so we obtain for the splitting 
\begin{subequations}
    \begin{align}
        \delta \hat{A}_{0} &= \delta A_0 - i\,(\beta \, k^i)^{\prime} \, A_i - A_i (\epsilon^i)^{\prime} \,,\\[4pt]
        \delta \hat{A}_{i} &= \delta A_i + k_i\, (\vec{k}\cdot \vec{A})\, \beta - i\, k_i\, (\vec{\epsilon}\cdot \vec{A}) - \dot{A}_i \alpha\,.
    \end{align}
\end{subequations}

\section{Synchronous gauge} \label{apendix:synch_gauge}
The sources of the vector equations of motion in synchronous gauge (Eq.~\ref{eq_mov_delta_A_synch}) to linear order in the metric shear are given by
\begin{subequations}
\begin{align}
    S_L &= \frac{1}{2} (A_L^{\prime} + \sigma_{\parallel} A_L + 2 \sigma_{v_1} A_{t_1}) \left(h^{\prime} + 8\eta^{\prime} \right) \nonumber\\[3pt]
    &- 2 \sigma_{v_1}^{\prime} \delta A_{t_1} - 2 \sigma_{v_1} \delta A_{t_1}^{\prime} - 2\mathcal{H}^{-2}\big[ (\mathcal{H}\sigma_{\parallel}^{\prime} - \mathcal{H}^{\prime} \sigma_{\parallel}) A_L^{\prime} \nonumber\\[3pt]
    &+ (\mathcal{H}\sigma_{v_1}^{\prime} - \mathcal{H}^{\prime} \sigma_{v_1}) A_{t_1}^{\prime} \big]\,\eta - 2 \frac{\sigma_{v_1}}{\mathcal{H}} A_{t_1}^{\prime} \eta^{\prime}\,,    
\end{align}
\begin{align}
    S_{t_1} &= -\frac{1}{2}\big[ A_t^{\prime} + (\sigma_{+} - \frac{1}{2}\sigma_{\parallel}) A_{t_1} \big] (h^{\prime} + 4 \eta^{\prime}) \nonumber \\[3pt]
    &- \mathcal{H}^{-2}\big[ A_{t_1}^{\prime} \big( \mathcal{H}^{\prime} (\sigma_{\parallel} - 2 \sigma_{+}) - \mathcal{H} (\sigma_{\parallel}^{\prime} - 2 \sigma_{+}^{\prime}) \big) \nonumber\\[3pt]
    &+ 2 A_L^{\prime} \left( \mathcal{H}\sigma_{v_1}^{\prime} + (6 \mathcal{H}^2 - \mathcal{H}^{\prime}) \sigma_{v_1} \right) \big] \eta+ 2 \sigma_{v_1} \delta A_L^{\prime} \nonumber\\[3pt]
    &- 2 \sigma_{v_1} A_L^{\prime}\, h  - 2 \frac{\sigma_{v_1}}{\mathcal{H}} A_{L}^{\prime} \eta^{\prime}\,.
\end{align}
\label{sources_synch}
\end{subequations}
We can neglect the terms containing the shear to linear order in the fluid variables.
The vector fluid variables at linear order in the metric shear are
\begin{subequations} 
\begin{align}
    \delta \rho_A  &= \frac{1}{a^4} \bigg[(A^{\prime}_L + A_L \sigma_{\parallel}+ 2\sigma_{v_1}A_{t_1}) \left(\delta A^{\prime}_L + \sigma_{\parallel}\,\delta A_L \right. \nonumber\\
    &\left. - i\,k\,\delta A_0\right) +  (A_{t_1}^{\prime} + (\sigma_+ - \frac{\sigma_{\parallel}}{2})A_{t_1}) \delta A_{t_1}^{\prime} + 2\rho_T a^4 \,\eta   \nonumber\\
    & + m^2 a^2 (A_L \,\delta A_L + A_{t_1} \,\delta A_{t_1})     + \left((\sigma_+ - \frac{\sigma_{\parallel}}{2}) A_{t_1}^{\prime}\right.\nonumber\\&+\left.2\sigma_{v_1}A_L^{\prime}\right) \delta A_{t_1} \bigg] - (\rho_L + \frac{2}{a^4}\sigma_{v_1}A_L^{\prime} A_{t_1}) ( h + 4 \eta )\,,\label{delta_rho_synch} 
    \end{align}
\begin{align} 
    \delta P_A &= \frac{1}{3a^4} \bigg[(A^{\prime}_L + A_L \sigma_{\parallel} + 2\sigma_{v_1} A_{t_1})\left(\delta A^{\prime}_L + \sigma_{\parallel}\,\delta A_L \right. \nonumber\\
    &- i\,k\,\delta A_0)  + (A_{t_1}^{\prime} + (\sigma_+ - \frac{\sigma_{\parallel}}{2})A_{t_1}) \delta A_{t_1}^{\prime}+ 6a^4P_T \, \eta  \nonumber   \\ 
    &- m^2 a^2 (A_L \,\delta A_L + A_{t_1} \,\delta A_{t_1}) + \left((\sigma_+ - \frac{\sigma_{\parallel}}{2}) A_{t_1}^{\prime}\right.\nonumber\\&+\left. 2\sigma_{v_1}A_L^{\prime} \right) \delta A_{t_1} \bigg]  - (P_L + \frac{2}{3a^4} \sigma_{v_1} A_L^{\prime} A_{t_1}) ( h + 4 \eta )\,,   \label{delta_P_synch} 
    \end{align}
\begin{align} 
    (\rho_A &+ P_A) \theta_A = \frac{k^2}{a^4} \left[A^{\prime}_T + A_T ( \sigma_+ - \frac{\sigma_{\parallel}}{2})\right] \delta A_{T_1}\nonumber\\
    &-i \frac{m^2 \, k}{a^2} \, A_L\, \delta A_0\,, 
    \end{align}
\begin{align} 
    (\rho_A &+ P_A) \delta\Sigma_{A\parallel}  = \frac{4}{3 a^4} \bigg[(A^{\prime}_L + A_L \sigma_{\parallel} + 2\sigma_{v_1} A_{t_1})\nonumber\\
    &\times(\delta A^{\prime}_L + \sigma_{\parallel}\,\delta A_L - i\,k\,\delta A_0)  - 3 a^4 P_L (h + 4 \eta) \nonumber\\
    &+ m^2 a^2 (A_{t_1} \,\delta A_{t_1} - A_L \,\delta A_L)  - \frac{\delta A^{\prime}_{t_1} }{2}  \left((\sigma_+ - \frac{\sigma_{\parallel}}{2})A_{t_1} \right.\nonumber\\
    &+\left.A_{t_1}^{\prime} \right)+ \delta A_{t_1}\big((\sigma_+ - \frac{\sigma_{\parallel}}{2}) \frac{A_{t_1}^{\prime}}{2} + 2\sigma_{v_1}A_L^{\prime} \big)   \bigg] \nonumber\\
    &- 4 P_T \, \eta-   \frac{8}{3a^4} \sigma_{v_1}A_L^{\prime}A_{t_1}(h + 4 \eta)\,,  \label{delta_shear_synch}
\end{align}
\end{subequations}
\noindent where we defined $\rho_L = \rho_A \big|_{A_t = 0}$, $\rho_T = \rho_A \big|_{A_L = 0}$, $P_L = P_A \big|_{A_t = 0}$ and $P_T = P_A \big|_{A_L = 0}$.
 
\section{Newtonian gauge} \label{apendix:newton_gauge}

In this Appendix we present the set of equations in Newtonian gauge as defined in Appendix \ref{appendix:gauge_change},

\begin{subequations}
\begin{align}
    \delta g_{00} &= -2 a^2 \phi\\[3pt]
    \delta g_{0i} &= 0 \\[3pt]
    \delta g_{ij} &= -2 a^2 (\gamma_{ij} + \frac{\sigma_{ij}}{\mathcal{H}}) \psi\,,
\end{align}
\end{subequations}

\noindent where we have neglected vector and tensor perturbations.
In Newtonian gauge, the vector  equations of motion are given by (\ref{eq_mov_delta_A_synch}), where the sources to linear order in the metric shear are given by
 
\begin{subequations}
\begin{align}
    S_L &= - 2 m^2 a^2 A_{L} \phi - 2 \sigma_{v_1} \delta A_{t_1}^{\prime} - 2 \sigma_{v_1}^{\prime} \delta A_{t_1} \\[3pt] 
    &+2\mathcal{H}^{-1} \left[ \frac{\mathcal{H^{\prime}}}{\mathcal{H}} (A_L^{\prime} \sigma_{\parallel} + A_t^{\prime} \sigma_{v_1}) - (\sigma_{\parallel}^{\prime} A_L^{\prime} + \sigma_{v_1}^{\prime} A_t^{\prime}) \right] \psi \nonumber\\[3pt]
    &+ (A_L^{\prime} + \sigma_{\parallel} A_L + 2 \sigma_{v_1} A_{t_1}) \left(\phi^{\prime} + \psi^{\prime} \right) - 2\frac{\sigma_{v_1}}{\mathcal{H}} A_{t_1}^{\prime} \psi^{\prime}\,,\nonumber
    \end{align}
    \begin{align} 
    S_{t_1}  &= - 2 m^2 a^2 A_t \phi - \mathcal{H}^{-2} \big[ A_t^{\prime}(\mathcal{H}^{\prime} (\sigma_{\parallel} - 2 \sigma_+)\\[3pt] 
    &- \mathcal{H} (\sigma_{\parallel}^{\prime} - 2 \sigma_+^{\prime})) + 2 A_L^{\prime} (\mathcal{H} \sigma_{v_1}^{\prime} - \mathcal{H}^{\prime} \sigma_{v_1}) \big] \psi \nonumber\\
    &+ (A_{t_1}^{\prime} + (\sigma_+ - \frac{\sigma_{\parallel}}{2}) A_t) (\phi^{\prime} + \psi^{\prime}) + 2 \sigma_{v_1} \delta A_L^{\prime}\nonumber\\
    &- 2\frac{\sigma_{v_1}}{\mathcal{H}} A_{L}^{\prime} \psi^{\prime}\nonumber\,.
\end{align}
\label{sources_newt}
\end{subequations} 

\vspace{-0.35cm}
The temporal constraint equation is given by
\begin{align}
    &\delta A_0 = - i \, \frac{k}{m^2 a^2 + k^2} \, \big[\delta A_L^{\prime}+ \sigma_{\parallel} \,\delta A_L + 2 \sigma_{v_1} \delta A_{t_1}\nonumber \\[5pt]
    &+ 2 \frac{\sigma_{v_1}}{\mathcal{H}} A_{t_1}^{\prime} \psi - (A_L^{\prime} + \sigma_{\parallel} A_L + 2 \sigma_{v_1} A_{t_1}) (\phi + \psi) \big] .
\end{align}

The VFDM fluid variables can be expressed in terms of the field and metric potentials as
\begin{subequations}
\begin{align}
    \delta &\rho_A = \frac{1}{a^4} \big[(A^{\prime}_L + A_L \sigma_{\parallel}+2\sigma_{v_1}A_{t_1})\left(\delta A^{\prime}_L + \sigma_{\parallel}\,\delta A_L \right.\nonumber\\& \left.- i\,k\,\delta A_0\right)  +  (A_{t_1}^{\prime} + (\sigma_+ - \frac{\sigma_{\parallel}}{2})A_{t_1}) \delta A_{t_1}^{\prime}  + m^2 a^2 \left(A_L \,\delta A_L\right.\nonumber\\ &\left.+ A_{t_1} \,\delta A_{t_1}\right) 
   + \left((\sigma_+ - \frac{\sigma_{\parallel}}{2}) A_{t_1}^{\prime} + 2 \sigma_{v_1} A_L^{\prime} \right)\delta A_{t_1}  \big] + 2\rho_A \, \psi  \nonumber\\
    &- (\rho_A + 3 P_A) \phi \,,\end{align}
\begin{align}
    \delta &P_A = \frac{1}{3 a^4} \big[(A^{\prime}_L + A_L \sigma_{\parallel} +2\sigma_{v_1}A_{t_1})\left(\delta A^{\prime}_L + \sigma_{\parallel}\,\delta A_L \right.\nonumber\\
    &\left.- i\,k\,\delta A_0\right) + \delta A_{t_1}^{\prime} (A_{t_1}^{\prime} + (\sigma_+ - \frac{\sigma_{\parallel}}{2})A_{t_1}) - m^2 a^2 \left(A_L \,\delta A_L \right.\nonumber\\&\left.+ A_{t_1} \,\delta A_{t_1}\right) +  \left( (\sigma_+ - \frac{\sigma_{\parallel}}{2})A_{t_1}^{\prime} + 2 \sigma_{v_1} A_L^{\prime} \right) \delta A_{t_1} \big] + 2P_A \psi\nonumber\\
    &- \frac{1}{3}(\rho_A + 3 P_A) \phi  \,,\end{align}
\begin{align} 
    (\rho_A &+ P_A) \theta_A =  + \frac{k^2}{a^4} \left[A^{\prime}_T + A_T ( \sigma_+ - \frac{\sigma_{\parallel}}{2})\right] \delta A_{T_1}\nonumber\\
    &-i \frac{m^2 \, k}{a^2} \, A_L\, \delta A_0\,,\end{align}
\begin{align}
   (\rho_A &+ P_A) \delta\Sigma_{A\parallel}  = \frac{4}{3 a^4} \big[(A^{\prime}_L + A_L \sigma_{\parallel} + 2 \sigma_{v_1} A_{t_1})\nonumber\\
   &\times(\delta A^{\prime}_L + \sigma_{\parallel}\,\delta A_L - i\,k\,\delta A_0) - 4 \sigma_{v_1} A_{t_1} A_L^{\prime} \phi\nonumber\\
   &- \frac{1}{2} (A_{t_1}^{\prime} + (\sigma_+ - \frac{\sigma_{\parallel}}{2})A_{t_1}) \delta A_{t_1}^{\prime} + \frac{1}{2}m^2 a^2 \big(A_{t_1} \,\delta A_{t_1}\nonumber  \\
    &- 2 A_L \,\delta A_L\big)    - \frac{1}{2} \left(A_{t_1}^{\prime}  (\sigma_+ - \frac{\sigma_{\parallel}}{2}) - 4\sigma_{v_1} A_L^{\prime}\right) \delta A_{t_1} \big]  \nonumber\\
    &+ 2 \,\Sigma_{\parallel}\,\psi + \frac{2}{3}\big[(\rho_T + 3 P_T) - 2(\rho_L + 3 P_L)\big] \phi  \,, 
\end{align}
\end{subequations}
\noindent where the fluid variables are defined as in the synchronous gauge (see Eqs. \ref{fluid_variables_definitions}). 

The Einstein's equations to linear order are\footnote{ We have checked that our  equations in Newtonian gauge reduces to  the ones derived in \cite{Pereira_2007} at linear order in the metric shear when the  assumption of no anisotropic shear for the matter sector is made.} 

\begin{subequations} 
\begin{equation}
    k^2 \psi + 3 \mathcal{H} (\psi^{\prime} + \mathcal{H} \phi) = - \frac{a^2}{2 m_P^2} \delta \rho 
    \label{Einstein00Newtonian} \,,
\end{equation}
\begin{align}    
    &k^2\, (\psi^{\prime} + \mathcal{H}\phi) - \frac{k^2}{2}  \mathcal{H}^{-2}\left[ (\mathcal{H}^{\prime} - 3 \mathcal{H}^2) \sigma_{\parallel}  \right.\nonumber\\
     &\left. - \mathcal{H} \sigma_{\parallel}^{\prime} \right]\psi= \frac{a^2}{2\,m_P^2}\, (\rho + P) \theta\,,
    \label{Einstein0iNewtonian}
\end{align}
\begin{align}
    &\psi^{\prime \prime} + (\mathcal{H}^2 + 2 \mathcal{H}^{\prime}) \phi + \mathcal{H} (\phi^{\prime} + 2 \psi^{\prime}) \nonumber\\
    &+ \frac{k^2}{3} (\psi - \phi) = \frac{a^2}{2 m_P^2} \delta P \,,
   \label{EinsteintraceNewtonian} 
\end{align}
\begin{align}
    &k^2 (\psi - \phi)  + 3 \left[\frac{\sigma_{\parallel}^{\prime}}{\mathcal{H}} +  \left(\frac{5}{2} - \frac{\mathcal{H}^{\prime}}{\mathcal{H}^2}\right) \sigma_{\parallel} \right] \psi^{\prime} \nonumber\\&+ \frac{3}{2} \sigma_{\parallel} \phi^{\prime}+ \frac{3}{2} \frac{\sigma_{\parallel}}{\mathcal{H}} \psi^{\prime \prime}+ \frac{3}{2}\left[ \frac{\sigma_{\parallel}^{\prime \prime}}{\mathcal{H}} + 2 \sigma_{\parallel}^{\prime} \left( 1 - \frac{\mathcal{H}^{\prime}}{\mathcal{H}^2} \right) \right.\nonumber\\
    &\left. + \sigma_{\parallel} \left( 2\frac{\mathcal{H}^{\prime}}{\mathcal{H}} \left(\frac{\mathcal{H}^{\prime}}{\mathcal{H}^2} - 1\right) - \frac{\mathcal{H}^{\prime\prime}}{\mathcal{H}^2} - \frac{k^2}{3} \right) \right] \psi \nonumber\\&
    + 3( \sigma_{\parallel}^{\prime} + 2 \mathcal{H} \sigma_{\parallel}) \phi= \frac{3 a^2}{2 m_P^2} (\rho + P) \delta\Sigma_{\parallel}  \label{EinsteinparNewtonian} 
\end{align}\label{einstein_eqs_new}
\end{subequations}

To recover the set of equations in synchronous gauge the vector perturbations has to be taken into account, since they mix with the scalar sector because of the anisotropy in the background. This checks has been performed, but the vector perturbations has been left out of this work for simplicity.

Using only that $\sigma_{\parallel}\ll\mathcal{H}$ it is immediate to see that the contribution of the metric shear $\sigma_{\parallel}$ can be neglected on the left-hand side of Eqs.~(\ref{Einstein00Newtonian}), (\ref{Einstein0iNewtonian}) and (\ref{EinsteintraceNewtonian}), but not in Eq.~(\ref{EinsteinparNewtonian}). Indeed, in the absence of nonlocal contributions, only  terms with the metric shear on the right-hand side of Eq.~(\ref{EinsteinparNewtonian})    do not vanish as $k\to 0$. Therefore,   we can see the metric shear contribution to the left-hand side of Eq.~(\ref{EinsteinparNewtonian}) will become important for small enough values of $k^2$. 

As in the synchronous gauge, in the $\rm{\Lambda CDM}$ model the initial conditions  are imposed far outside the horizon and deep in radiation era, thus obtaining \cite{Ma_1995}
\begin{subequations}
\begin{align}
    \theta_{\nu} = \theta_{\gamma}  &= \frac{1}{2} k^2 \tau \,\phi \,,\\
    \phi &= \frac{20 \,C}{15+ 4 R_{\nu}}\,,  \\
    \psi &= (1+\frac{2}{5} R_{\nu})\phi\,,
\end{align}
\end{subequations} 
\noindent where $C$ is a constant and $R_{\nu} = \frac{\rho_{\nu}}{\rho_{\gamma}}$. In particular, we see that $\psi$ and $\phi$ are of the same order at early times.

In the presence of the VFDM, 
the solutions for the vector equations of motion at early times can be written as a power law $\delta A \propto \tau^{\alpha}$, and we can see that in a radiation dominated universe we can neglect the terms with the metric shear in Eq.~ (\ref{eq_mov_delta_A_synch}), with sources given in Eq.~ (\ref{sources_newt}).
The constants of integration are chosen by imposing adiabatic initial conditions for the VFDM component ($\delta\rho_A/\rho_A'=\delta\rho_\gamma/\rho_\gamma'$), thus obtaining
\begin{equation}\label{AAM}
    \delta \vec{A} = - \vec{A}  \,\psi   \,.
\end{equation} 
 
With this asymptotic initial condition for the field we can  calculate the other initial expressions for the fluid variables in Newtonian gauge,
\begin{align}
    \delta P_A &= -\frac{2}{3} \rho_A \phi \,,\\[3pt]
    (\rho_A+P_A)\delta\Sigma_{\parallel, A} &= \frac{2}{3}\frac{-2c_L^2 + c_{t_1}^2}{a^4}\,\phi\,,\label{vecshearNewt}\\[3pt]
    (\rho_A+P_A)\theta_A &= \frac{c_L^2 (\phi + 2 \psi) - c_{t_1}^2\psi}{a^4}\, k^2\tau\,.
\end{align}
This adiabatic initial conditions gives a nonzero contributions for the vector  shear far outside the horizon. However, as can be seen by inserting Eq.~(\ref{vecshearNewt})   into  Eq.~(\ref{EinsteinparNewtonian})  and using (\ref{sigmaparalel}) and the radiation domination evolution for the scale factor,   we can see this contribution  cancels the one with the metric shear.

Therefore, as in the synchronous gauge, the large infrared contribution of the vector field is exactly canceled by the contribution of the metric shear, so we recover $\rm{\Lambda CDM}$ behavior in the metric perturbations and the rest of the species at early times.\\

\section{Adiabatic mode for the vector field and the Weinberg's construction  }  \label{apendix:WeinbergAD}

In \cite{Weinberg_2003} the author uses a residual symmetry to find exact solutions to the system for modes far outside the horizon. In   $\rm{\Lambda CDM}$ the linearized Einstein's equations in Newton gauge are invariant under a redefinition of the time coordinates and a rescaling of spatial coordinates for $k=0$, namely 
\begin{equation}
    \begin{cases}
        t \to t + \epsilon(t) \\
        x \to x(1-\lambda)
    \end{cases},
\end{equation}where $t$ is the cosmic time $dt=a \,d\tau$.

In the case of VFDM, the presence of anisotropies in the early Universe could significantly break this invariance. However, as shown in the previous Appendix,  in the  equations that do not vanish  for $k= 0$ in  $\rm{\Lambda CDM}$ the shear (and the backreaction of the VFDM  in radiation domination era) can be neglected  and, therefore, those equations reduce to the standard ones in  $\rm{\Lambda CDM}$.   Then, by performing the transformation of the metric and each species, we can find the  corresponding solutions for the system at $k=0$:

\begin{subequations}\label{WAMALL}
\begin{align}
    \delta A_i &=   - \epsilon   \dot{A_i}+ \lambda  A_i\,, \label{WAM}\\[3pt]
    \phi &= - \dot{\epsilon}\,, \\[3pt]
    \psi &= H \, \epsilon - \lambda\,,
\end{align}
\end{subequations}

\noindent where  a dot stands for a derivative with respect to cosmic time. Since, in the radiation domination era,  the background field satisfies  $\dot{A_i} \simeq {H}A_i$, Eq.~(\ref{WAM})  reduces to Eq.~(\ref{AAM}).  As shown in the Appendix above, despite the fact that both the right-hand side and the left-hand side of the nondiagonal $ij$ Einstein equations do not vanish separately in the limit $k\to 0$, as happens for $\rm{\Lambda CDM}$,  the two sides are equal. Hence the equations that in $\rm{\Lambda CDM}$ vanish in the limit $k\to0$ are also satisfied and, therefore, the above solution  is a physical solution that can be extended to  $k\neq0$  for $k\to0$. More generally, one can see that the solution in (\ref{WAM}) satisfies the VFDM equation of motion at leading order in the shear  for $k\to0$, not only in radiation dominated era but also afterwards,  provided that  $\lambda$ is constant \cite{Ma_1995},  by replacing (\ref{WAMALL}) into  Eq.~(\ref{eq_mov_delta_A_synch}) (with the sources given in  Eq.~(\ref{sources_newt})) and using the equation for the background field.

\bibliographystyle{ieeetr}
\bibliography{reference}

\begin{thebibliography}{10}

\bibitem{Marsh_2016}
D.~J. Marsh, ``Axion cosmology,'' {\em Phys. Rep.}, vol.~643, pp.~1--79, jul 2016.

\bibitem{Ferreira_2021}
E.~G.~M. Ferreira, ``Ultra-light dark matter,'' {\em The Astronomy and Astrophysics Review}, vol.~29, sep 2021.

\bibitem{Schive:2014dra}
H.-Y. Schive, T.~Chiueh, and T.~Broadhurst, ``{Cosmic Structure as the Quantum Interference of a Coherent Dark Wave},'' {\em Nature Phys.}, vol.~10, pp.~496--499, 2014.

\bibitem{Hlozek:2014lca}
R.~Hlozek, D.~Grin, D.~J.~E. Marsh, and P.~G. Ferreira, ``{A search for ultralight axions using precision cosmological data},'' {\em Phys. Rev. D}, vol.~91, no.~10, p.~103512, 2015.

\bibitem{Ure_a_L_pez_2016}
L.~A. Ure{\~{n} }a-L{\'{o}}pez and A.~X. Gonzalez-Morales, ``Towards accurate cosmological predictions for rapidly oscillating scalar fields as dark matter,'' {\em Journal of Cosmology and Astroparticle Physics}, vol.~2016, pp.~048--048, jul 2016.

\bibitem{Pk_linear_lyman_1}
V.~Ir\v{s}i\v{c}, M.~Viel, M.~G. Haehnelt, J.~S. Bolton, and G.~D. Becker, ``{First constraints on fuzzy dark matter from Lyman-$\alpha$ forest data and hydrodynamical simulations},'' {\em Phys. Rev. Lett.}, vol.~119, no.~3, p.~031302, 2017.

\bibitem{Rogers:2020ltq}
K.~K. Rogers and H.~V. Peiris, ``{Strong Bound on Canonical Ultralight Axion Dark Matter from the Lyman-Alpha Forest},'' {\em Phys. Rev. Lett.}, vol.~126, no.~7, p.~071302, 2021.

\bibitem{Kobayashi:2017jcf}
T.~Kobayashi, R.~Murgia, A.~De~Simone, V.~Ir\v{s}i\v{c}, and M.~Viel, ``{Lyman-$\alpha$ constraints on ultralight scalar dark matter: Implications for the early and late universe},'' {\em Phys. Rev. D}, vol.~96, no.~12, p.~123514, 2017.

\bibitem{Hlozek:2017zzf}
R.~Hlozek, D.~J.~E. Marsh, and D.~Grin, ``{Using the Full Power of the Cosmic Microwave Background to Probe Axion Dark Matter},'' {\em Mon. Not. Roy. Astron. Soc.}, vol.~476, no.~3, pp.~3063--3085, 2018.

\bibitem{Lague:2021frh}
A.~Lagu\"e, J.~R. Bond, R.~Hlo\v{z}ek, K.~K. Rogers, D.~J.~E. Marsh, and D.~Grin, ``{Constraining ultralight axions with galaxy surveys},'' {\em JCAP}, vol.~01, no.~01, p.~049, 2022.

\bibitem{Dentler:2021zij}
M.~Dentler, D.~J.~E. Marsh, R.~Hlo\v{z}ek, A.~Lagu\"e, K.~K. Rogers, and D.~Grin, ``{Fuzzy dark matter and the Dark Energy Survey Year 1 data},'' {\em Mon. Not. Roy. Astron. Soc.}, vol.~515, no.~4, pp.~5646--5664, 2022.

\bibitem{lin2023constraining}
H.~Lin, F.~Deng, Y.~Gong, and X.~Chen, ``Constraining ultralight axions with csst weak gravitational lensing and galaxy clustering photometric surveys,'' 2023.

\bibitem{Lague:2023wes}
A.~Lagu\"e, B.~Schwabe, R.~Hlo\v{z}ek, D.~J.~E. Marsh, and K.~K. Rogers, ``{Cosmological simulations of mixed ultralight dark matter},'' 10 2023.

\bibitem{Arias:2012az}
P.~Arias, D.~Cadamuro, M.~Goodsell, J.~Jaeckel, J.~Redondo, and A.~Ringwald, ``{WISPy Cold Dark Matter},'' {\em JCAP}, vol.~06, p.~013, 2012.

\bibitem{Graham_2016}
P.~W. Graham, J.~Mardon, and S.~Rajendran, ``Vector dark matter from inflationary fluctuations,'' {\em Physical Review D}, vol.~93, may 2016.

\bibitem{Agrawal:2018vin}
P.~Agrawal, N.~Kitajima, M.~Reece, T.~Sekiguchi, and F.~Takahashi, ``{Relic Abundance of Dark Photon Dark Matter},'' {\em Phys. Lett. B}, vol.~801, p.~135136, 2020.

\bibitem{Dror:2018pdh}
J.~A. Dror, K.~Harigaya, and V.~Narayan, ``{Parametric Resonance Production of Ultralight Vector Dark Matter},'' {\em Phys. Rev. D}, vol.~99, no.~3, p.~035036, 2019.

\bibitem{Co:2018lka}
R.~T. Co, A.~Pierce, Z.~Zhang, and Y.~Zhao, ``{Dark Photon Dark Matter Produced by Axion Oscillations},'' {\em Phys. Rev. D}, vol.~99, no.~7, p.~075002, 2019.

\bibitem{Long:2019lwl}
A.~J. Long and L.-T. Wang, ``{Dark Photon Dark Matter from a Network of Cosmic Strings},'' {\em Phys. Rev. D}, vol.~99, no.~6, p.~063529, 2019.

\bibitem{Nakayama:2019rhg}
K.~Nakayama, ``{Vector Coherent Oscillation Dark Matter},'' {\em JCAP}, vol.~10, p.~019, 2019.

\bibitem{Kaneta:2023lki}
K.~Kaneta, H.-S. Lee, J.~Lee, and J.~Yi, ``{Misalignment mechanism for a mass-varying vector boson},'' {\em JCAP}, vol.~09, p.~017, 2023.

\bibitem{Kitajima:2023fun}
N.~Kitajima and K.~Nakayama, ``{Viable vector coherent oscillation dark~matter},'' {\em JCAP}, vol.~07, p.~014, 2023.

\bibitem{Cembranos_2012_isotropy_theorem}
J.~A.~R. Cembranos, C.~Hallabrin, A.~L. Maroto, and S.~J.~N. Jare{\~{n}}o, ``Isotropy theorem for cosmological vector fields,'' {\em Physical Review D}, vol.~86, jul 2012.

\bibitem{Golovnev_2008}
A.~Golovnev, V.~Mukhanov, and V.~Vanchurin, ``Vector inflation,'' {\em Journal of Cosmology and Astroparticle Physics}, vol.~2008, p.~009, June 2008.

\bibitem{Weinberg_2003}
S.~Weinberg, ``Adiabatic modes in cosmology,'' {\em Physical Review D}, vol.~67, jun 2003.

\bibitem{Pereira_2007}
T.~S. Pereira, C.~Pitrou, and J.-P. Uzan, ``Theory of cosmological perturbations in an anisotropic universe,'' {\em Journal of Cosmology and Astroparticle Physics}, vol.~2007, pp.~006--006, sep 2007.

\bibitem{Ford_vector_inflation}
L.~H. Ford, ``Inflation driven by a vector field,'' {\em Phys. Rev. D}, vol.~40, pp.~967--972, Aug 1989.

\bibitem{Planck:2018vyg}
N.~Aghanim {\em et~al.}, ``{Planck 2018 results. VI. Cosmological parameters},'' {\em Astron. Astrophys.}, vol.~641, p.~A6, 2020.
\newblock [Erratum: Astron.Astrophys. 652, C4 (2021)].

\bibitem{Akarsu_2019}
Özgür Akarsu, S.~Kumar, S.~Sharma, and L.~Tedesco, ``Constraints on a bianchi type i spacetime extension of the standard $\rm{\Lambda}\rm{CDM}$ model,'' {\em Physical Review D}, vol.~100, jul 2019.

\bibitem{paper_in_prep}
T.~Ferreira~Chase, M.~Leizerovich, D.~L\'{o}pez~Nacir, and S.~Landau, ``{Cosmological perturbations with ultralight vector dark matter fields: numerical implementation in CLASS (to appear)},''

\bibitem{Pitrou:2013hga}
C.~Pitrou, X.~Roy, and O.~Umeh, ``{xPand: An algorithm for perturbing homogeneous cosmologies},'' {\em Class. Quant. Grav.}, vol.~30, p.~165002, 2013.

\bibitem{xact}
J.~M. Mart\'in-Garc\'ia {\em , http://www.xact.es}.

\bibitem{Ma_1995}
C.-P. Ma and E.~Bertschinger, ``Cosmological perturbation theory in the synchronous and conformal newtonian gauges,'' {\em The Astrophysical Journal}, vol.~455, p.~7, dec 1995.

\end{thebibliography}

\end{document}